# New CMOS Compatible Platforms for Integrated Nonlinear Optical Signal Processing

D. J. Moss[1,2] and R. Morandotti[2]

[1] School of Electrical and Computer Engineering, RMIT University, Melbourne, Victoria, 3001 Australia
[2] INRS-EMT, 1650 Boulevard Lionel Boulet, Varennes, Québec, Canada, J3X 1S2
dmoss@physics.usyd.edu.au

*Abstract*

Nonlinear photonic chips have succeeded in generating and processing signals all-optically with performance far superior to that possible electronically - particularly with respect to speed. Although silicon-on-insulator has been the leading platform for nonlinear optics, its high two-photon absorption at telecommunications wavelengths poses a fundamental limitation. This paper reviews some of the recent achievements in CMOS-compatible platforms for nonlinear optics, focusing on amorphous silicon and Hydex glass, highlighting their potential future impact as well as the challenges to achieving practical solutions for many key applications. These material systems have opened up many new capabilities such as on-chip optical frequency comb generation and ultrafast optical pulse generation and measurement. **Keywords-component; CMOS Silicon photonics, Integrated optics, Integrated optics Nonlinear; Integrated optics materials**

## I. INTRODUCTION

All-optical signal processing chips have been demonstrated in silicon nanowires [1] and chalcogenide glass (ChG) waveguides [2, 3] including all-optical logic [4], demultiplexing from 160Gb/s [5] to over 1Tb/s [6] via four-wave mixing (FWM), all-optical performance monitoring using slow light at bit rates up to 640Gb/s [7-9], optical regeneration [10,11], and many other processes. The third order nonlinear efficiency of all-optical devices can be improved dramatically by increasing the waveguide nonlinear parameter, $\gamma = \omega n_2 / c A_{eff}$ (where $A_{eff}$ is the waveguide effective area, $n_2$ the Kerr nonlinearity, and $\omega$ the pump frequency) as well as by using resonant structures to enhance the local field intensity. High index materials, such as semiconductors and ChG, offer excellent optical confinement and high values of $n_2$, a powerful combination that has produced extremely high values of $\gamma = 200,000$ W$^{-1}$ km$^{-1}$ for silicon nanowires [1], and 93,400 W$^{-1}$ km$^{-1}$ in ChG nanotapers [2]. Yet silicon suffers from high nonlinear losses due to two-photon absorption (TPA) and the resulting free carriers. Even if free carriers can be eliminated by using p-i-n junctions, its poor intrinsic nonlinear figure of merit (FOM = $n_2 / (\beta \lambda)$, where $\beta$ is the two-photon absorption coefficient) is very low. While this FOM is considerably higher for ChG, the development of fabrication processes for these newer materials is at a much earlier stage. The potential impact of a low FOM was dramatically illustrated in recent experiments in silicon at longer wavelengths below the TPA threshold [12,13]. While TPA can, in some instances, be turned around and used to advantage for all-optical functions [14-17], for the most part in the telecom band the low FOM of c-Si poses a fundamental limitation and is a material property that cannot be improved.

Recently, new platforms for nonlinear photonics based on CMOS compatible Hydex and silicon nitride [17-26] have been demonstrated that exhibit virtually no nonlinear absorption in the telecom band. These platforms have been the basis of many ground breaking demonstrations, and in particular have revolutionized the field of micro-resonator based optical frequency combs and ultrashort pulsed modelocked lasers [31], showing great promise for applications to ultrahigh bandwidth telecommunications [32]. More recently it has been shown to be capable of forming novel and efficient sources for quantum optical applications [33].

This paper reviews some of the progress made in Hydex glass [26] as well as amorphous silicon – a relatively new and promising nonlinear material optical platform. The first on-chip micro-resonator based optical parametric oscillators were demonstrated in 2010 [19, 34]. These papers showed that Kerr nonlinearity based frequency comb sources could be achieved in integrated optical chips in ring resonators with relatively modest Q-factors compared to the extremely high Q micro-toroid structures initially studied [35]. In Hydex glass, CW optical "hyper-parametric" oscillation in a micro-ring resonator with a Q factor of 1.2 million was demonstrated with a differential slope efficiency 7.4% for a single oscillating mode out of a single port, a CW threshold power as low as 50mW, and a controllable range of frequency spacing from 200GHz to more than 6THz. Subsequently, a stable modelocked laser based on this device was demonstrated with pulse repetition rates as high as 800GHz [21]. Further, novel functions in ultra-long (45cm) spiral waveguides, including a device capable of measuring both the amplitude and phase of ultrafast optical pulses, using an approach based on

Spectral Phase Interferometry by Direct Electric-field Reconstruction (SPIDER) [22] were demonstrated, as well as optical parametric gain approaching +20dB [25].

The success of these devices is due to their very low linear loss, a high nonlinearity parameter of $\gamma \cong 233 W^{-1} km^{-1}$ as well as negligible nonlinear losses up to extremely high intensities ($25 GW/cm^2$) [20]. The low loss, design flexibility, and CMOS compatibility of these devices will enable multiple wavelength sources for telecommunications, computing, metrology and other areas.

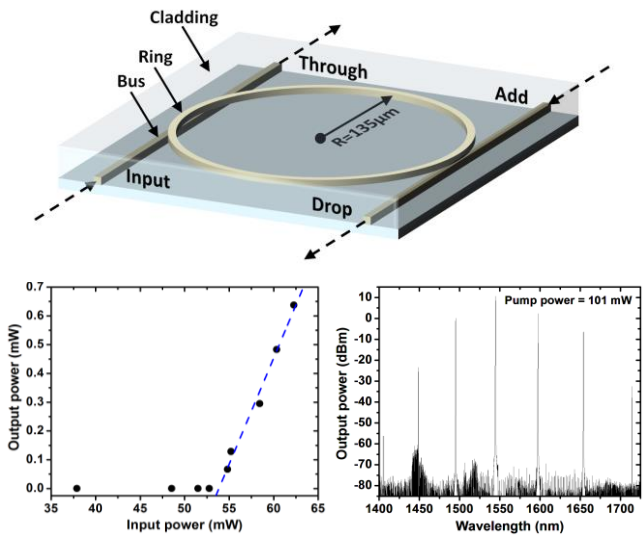

Figure 1. (top) Output spectra of hyperparametric oscillator near threshold (50mW) (top let) and at full pumping power (101mW, top right). Output spectrum when pumping closer to the zero dispersion point (bottom left) and output power of a single line, single port vs pump power.

## II. INTEGRATED HYPERPARAMETRIC OSCILLATOR

Figure 1 shows the device structure for the integrated oscillators - a four port micro-ring resonator with radius $\cong 135\mu m$ comprised of waveguides with a cross section of $1.45\mu m \times 1.5\mu m$. The bus waveguides used to couple light in and out of the resonator have the same cross section and are buried in $SiO_2$. The waveguide core is low loss Hydex glass with n=1.7 and a core-cladding contrast of 17% [19]. The films were deposited by standard chemical vapor deposition (PECVD) and device patterning and fabrication were performed by photolithography and reactive ion etching before over-coating with silica glass. Propagation losses were 0.04dB/cm and coupling loss to fiber pigtails of $\cong$ 1.5dB / facet. The ring resonator has a free spectral range (FSR) of 200GHz and a FWHM bandwidth of 1.3pm, corresponding to a Q factor of 1.2 million. The dispersion in these waveguides is shown in Figure 2, and is anomalous [17] over most of the C-band. The zero dispersion point for TM polarization is ~1560nm with $\lambda < 1560nm$ being anomalous and $\lambda > 1560nm$ normal.

Figure 1 shows the output spectra for a TM polarized pump beam at 1544.15nm with a pump power of 101mW. Initial oscillation occurs at 1596.98nm, 52.83nm away from the pump showing a wide frequency spacing of almost 53nm. The MI gain curve is shown in Figure 3, which peaks at ~1590nm, which agrees very well with the observed initial lasing wavelength. Figure 1 (bottom left) also shows the power of the mode at 1596.98nm exiting one port (drop) versus input pump power, showing a (single line) differential slope efficiency above threshold of 7.4%. The maximum total output power was at 101mW pump at 1544.15nm where we obtained 9mW in all modes out of both ports, with 2.6mW in a single line at 1596.98nm from a single port. The total oscillating mode power of 9mW represents a total conversion efficiency of 9%. When pumping at 1565.19 nm (normal dispersion) no oscillation was observed as expected. When pumping near zero dispersion, at 1558.65nm, we observed lasing with a spacing of 28.15nm, agreeing with the expected shift in the MI gain profile.

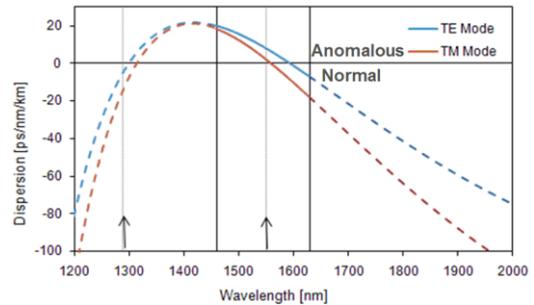

Figure 2. Dispersion curves for TE and TM polarizations in the waveguides. Solid curves are experimental data while dashed curves are theoretical curves.

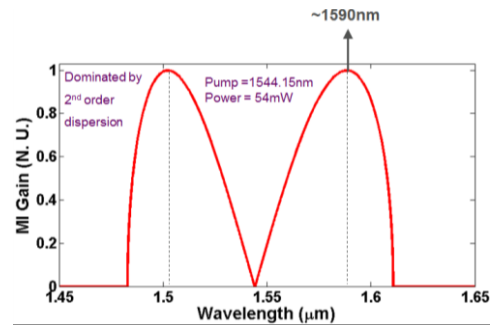

Figure 3. Calculated modulational instability gain for optical parametric oscillator.

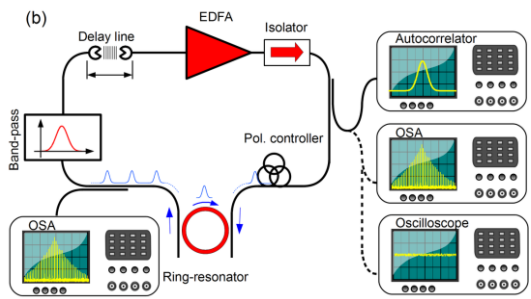

Figure 4. Experimental configuration for filter-driven dissipative four wave mixing based modelocked laser.

Since the initial demonstration of Kerr resonator based freqency comb generation [35] and the subsequent extension to comb generation in integrated waveguides [19, 34] there has been an explosion of activity in this area [36-40] with demonstrations of cavity solitons [31], applications to ultrahigh bandwidth tele-communications [32], ultrahigh speed arbitrary waveform generation [40] and many others. Nontheless, there has only been one approach proposed and demonstrated so far that has managed to achieve intrinsically stable modelocked operation based on a microresonator [21], and we turn to this next.

### III. SELF-LOCKED LASERS

Passively mode-locked lasers have generated the shortest optical pulses to date [41 - 53]. Many different approaches have been proposed to achieve very high and flexible repetition rates at frequencies well beyond active mode-locking, from very short laser cavities with large mode frequency spacings (large FSR) [21,41,45-48], where a very high repetition rate is achieved by simply reducing the pulse round-trip time, to schemes where multiple pulses are produced in each round trip [33,47,48]. In 1997 Yoshida et al. [48], introduced dissipative FWM [49, 51], where a Fabry Pérot filter is inserted in the main cavity to suppress all but a few periodically spaced modes, leading to a train of pulses with a very high repetition rate. Although dissipative FWM yielded transform limited pulses at very high repetition rates, a common problem is supermode instability where multiple pulses circulate in a cavity. This is a consequence of the much smaller cavity mode frequency spacings of a few megahertz or less, which allows many modes to oscillate within the Fabry Pérot filter bandwidth, which produces extremely unstable operation [52].

Figure 4 shows the configuration of the first mode-locked laser [21, 23] based on a nonlinear monolithic high-Q (quality factor) resonator that achieved extremely stable operation at high repetition rates while maintaining very narrow linewidths. The resonator is used as both filter and nonlinear element. This mode-locking scheme is termed filter-driven four-wave-mixing (FD-FWM). It operates in a way which is in stark contrast to traditional dissipative FWM schemes where the nonlinear interaction occurs in the fibre and is then filtered separately by a linear Fabry Pérot filter. The micro-ring resonator is embedded in an Erbium doped-fibre loop cavity containing a passband filter with a bandwidth large enough to pass all of the oscillating lines. A delay line controls the phase of the main cavity modes with respect to the ring modes.

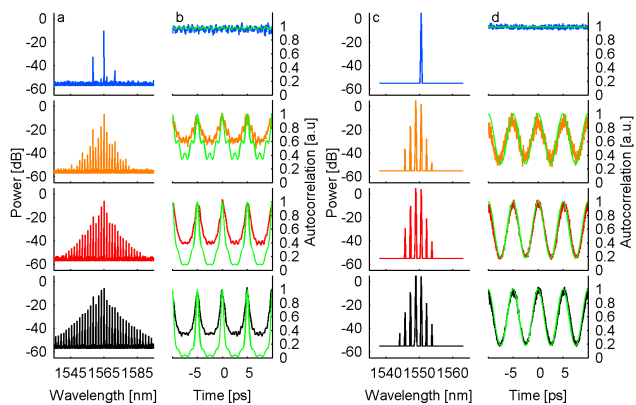

Figure 5. (left) Optical output for long cavity length (unstable) laser and (right0 output of short cavity (stable laser). Green curves are theoretical autocorrelation plots.

Figure 5 compares the optical output of a laser based on a 33m long cavity along with one that used a very short EYDFA (Erbium Ytterbium), with a fiber loop length of only 3m. The two configurations had significantly different main cavity lengths (3m and 33m) with different FSRs (68.5MHz and 6MHz) as well as different saturation powers. Figure 5 compares the optical spectra of the pulsed output along with the temporal traces obtained by an auto-correlator for the two systems at four pump powers. The pulses visible in the autocorrelation trains had a temporal duration that decreases noticeably as the input power increases, as expected for a typical passive mode-locking scheme.

From these plots it would appear that the long laser had better overall performance since its pulsewidth was shorter. However, the key issue of laser stability is better illustrated by a comparison between the experimental autocorrelation traces with the calculated traces (green) in Figure 5. While a perfect match is found for the short length EDFA case, the long cavity design shows a considerably higher background, thus clearly distinguishing unstable from stable laser operation. To quantify the pulse-to-pulse stability we recorded the electrical radio-frequency (RF) spectrum of the envelope signal, collected at the output using a fast photo-detector. Unstable oscillation was always observed [21] for the long cavity due to the presence of a large number of cavity modes oscillating in the ring resonance. In contrast, the short-cavity could easily be stabilized to give very stable operation by adjusting the main cavity length in order to center a single main cavity mode with respect to the ring resonance, thus eliminating any main cavity low-frequency beating. This self-locked approach has also been applied to CW operation where a two comb lines at each resonance were obtained to yield an ultra-pure radio frequency beat tone [23].

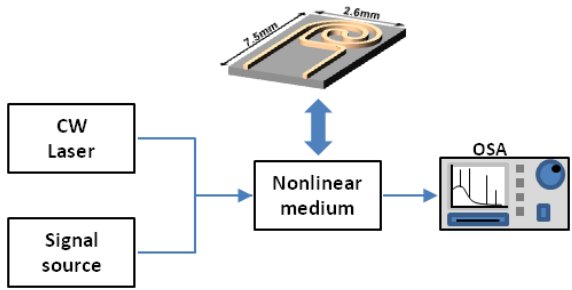

Figure 6. Experimental configuration for filter-driven dissipative four wave mixing based modelocked laser.

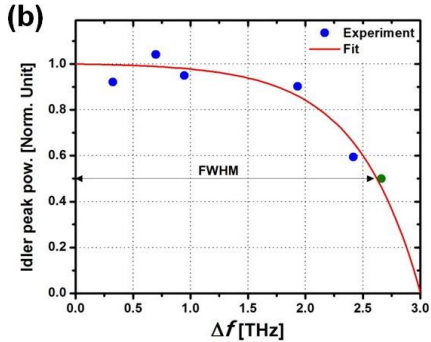

Figure 7. Experimental configuration for filter-driven dissipative four wave mixing based modelocked laser.

## IV. RADIO FREQUENCY SPECTRUM ANALYZER

Photonics offers the capability of generating and measuring ultrashort optical pulses, with bandwidths of many THz and at repetition rates of hundreds of GHz. Performing temporal diagnostics at these speeds is extremely difficult and yet essential to achieve high optical signal fidelity of fundamental noise parameters such as time jitter and amplitude noise, critical for achieving the maximum performance of many devices such as high frequency - clock optical modules [54, 55]. The traditional way of measuring the RF spectrum consists of recording the temporal intensity profile by an ultra-fast photo detector and then processing this signal, but this approach is limited to around 50 GHz.

A key breakthrough was the all-optical RF-spectrum analyzer, introduced by Dorrer and Maywar [56], based on optical mixing between a signal and CW probe via the Kerr ($n_2$) nonlinearity. In this approach, a single measurement of the CW probe optical spectrum with an optical spectrum analyzer (OSA) yields the intensity power spectrum of the signal under test. This approach can achieve much broader bandwidths than electronic methods, with a trade-off between sensitivity and bandwidth, or between the nonlinear response and total dispersion of the waveguide. Since then much progress has been made in realizing this device in integrated form [57 - 61]. Increasing the device length enhances the nonlinear response but results in increased dispersion that reduces the frequency response. For this reason, an optical integration platform with high nonlinearity and low net dispersion (waveguide plus material) is highly desirable. The first demonstration of an integrated all-optical RF spectrum analyzer [57] was achieved in chalcogenide waveguides in only a few centimeters of length. This was followed by a device on a silicon nanowire [59] and was subsequently used to monitor dispersion of ultrahigh bandwidth coherent signals [61].

Recently [60], we reported an integrated RF spectrum analyzer based on Hydex glass [26]. Figure 6 shows the device configuration of the RF spectrum analyzer while Figure 7 shows its measured frequency response showing a 3dB bandwidth of about 2.6THz, limited by our system measurement capability. We believe the intrinsic bandwidth is substantially higher than this since the simple theoretical prediction used in [52, 53] yields a bandwidth more than 100THz. In practice this would likely be limited by higher order dispersion, mode cutoff and even absorption bands, since the simple model of [56, 57] does not include these effects. Figure 8 shows the results of using this device to measure the RF frequency response of the ultrahigh repetition rate laser discussed in Section III that emitted sub picosecond pulses at repetition rates of 200 GHz and 400GHz. This device allowed us to analyze these lasers according to the noise burst model, which identifies very rapid intensity fluctuations of the laser pulses as the main source of noise [62, 63]. The RF spectra for both lasers show sensitivity to high frequency noise not detectable by other methods.

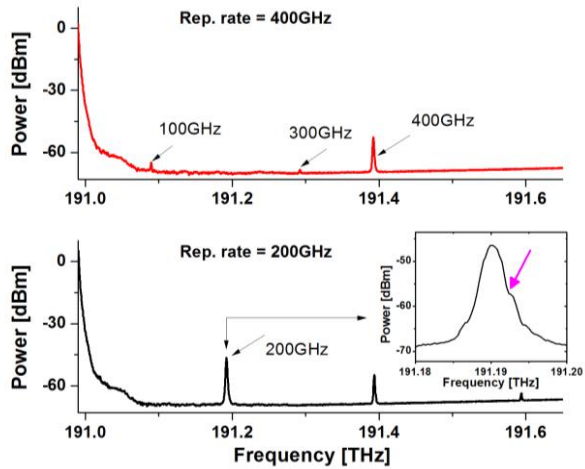

Figure 8. Measured RF spectrum of high frequency modelocked laser described in section III.

## V. AMORPHOUS SILICON

The ideal nonlinear optical platform would have all the attributes of silicon-on-insulator but with a FOM >1. Amorphous silicon, of interest as a nonlinear material for some time [64], was recently suggested [65] as a promising alternative to silicon for nonlinear optics, promising a larger FOM than c-Si [66, 67]. Since then much progress has been made [68 - 73]. Recent results have confirmed the possibility of increasing the FOM from around 1 [68, 69] to as high as 2 at telecommunication wavelengths [70], enabling high parametric gains of over +26dB over the C-band [69]. Table I compares the nonlinear properties of a range of nonlinear materials including Hydex, silicon nitride, and crystalline and amorphous silicon where it can be seen that a-Si shows a significant improvement in both Kerr nonlinearity and FOM over c-Si. Our recent results [71] also showed a significant enhancement in stability over previous results [69, 70]. Our measurements yielded both a record high nonlinear FOM of 5 – over 10 x SOI, and a nonlinearity ($\gamma$ factor) of almost 5x silicon. This may seem counterintuitive since Kramers - Kronig relations normally imply that increasing the bandgap to decrease nonlinear absorption decreases the nonlinear response. For silicon, however, the real part of the nonlinear susceptibility is largely determined by the direct transitions [74 - 78], while the TPA in the telecom band arises from indirect transitions. For a-Si therefore, it could be hypothesized that an increase in the *indirect* bandgap (reducing TPA) could be accompanied by a decrease in the *direct* bandgap (increasing the Kerr nonlinearity. Figure 10 illustrates the principle showing the indirect two-photon absorption in silicon as well as a full bandstructure of silicon from [79]. The direct bandgap in silicon, which is around 3.5eV, is largely responsible for determining the real part of the nonlinear response whereas the indirect bandgap dominates transitions involving multiphoton absorption. Therefore decreasing the direct bandgap will not significantly increase the two-photon absorption and conversely, increasing the indirect bandgap will not significantly affect $n_2$. Table I summarizes the nonlinearities of key nonlinear materials including silicon, silicon nitride, Hydex and amorphous silicon, where it can be seen that amorphous silicon has by far the best combination of parameters including high $n_2$ and simultaneously high FOM.

Finally, a key goal for all-optical chips is to reduce device footprint and operating power, and the dramatic improvement in the FOM of a-Si raises the possibility of using slow-light structures [7 - 9] to allow devices to operate at mW power levels with sub-millimeter lengths.

## VI. CONCLUSION

We review a wide range of on-chip devices based on a CMOS compatible Hydex glass platform and amorphous silicon. These devices have significant potential for applications requiring CMOS compatibility for both telecommunications and on-chip WDM optical interconnects for computing.

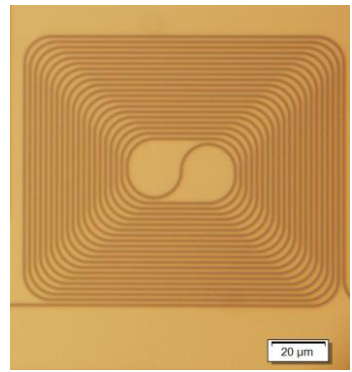

Figure 9. Amorphous Si waveguide spiral reported in Ref[71].

## Table I
Nonlinear parameters for key materials

|  | a-Si | c-Si | SiN | Hydex |
|---|---|---|---|---|
| $n_2$ (x fused silica[1]) | 700 | 175 | 10 | 5 |
| $\gamma$ [W$^{-1}$m$^{-1}$] | 1200 | 300 | 1.4 | 0.25 |
| $\beta_{TPA}$ [cm/GW] | 0.25 | 0.9 | 0 [2] | 0 [3] |
| FOM | 5 | 0.3 | ∞ | ∞ |

[1] $n_2$ for fused silica = 2.6 x 10$^{-20}$ m$^2$/W
[2] no nonlinear absorption has been observed in SiN nanowires.
[2] no nonlinear absorption has been observed in Hydex waveguides up to intensities of 25GW/cm$^2$.

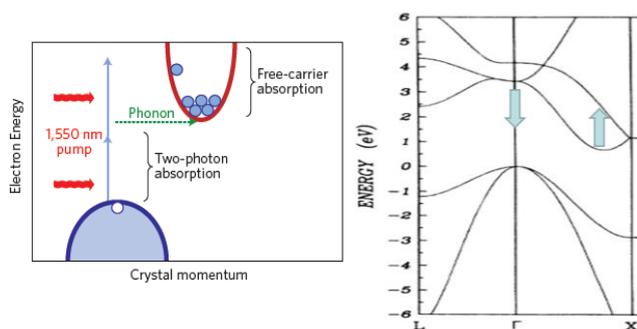

Figure 10. Band diagram of silicon showing the indirect two photon absorption in the telecom band from Ref [80]. Right: Bandstructure of silicon from Ref[79] showing the different effects of reducing the direct bandgap while increasing the indirect bandgap, as is likely similar to what occurs in amorphous silicon.